\documentclass{ws-ijmpcs}

\begin{document}


%
\catchline{}{}{}{}{}
%

\title{Electroweak measurements from $W$, $Z$ and photon final states}

\author{Hang Yin\\
(On behalf of ATLAS, CDF, CMS, D0, and LHCb collaborations)}

\address{ Fermi National Accelerator Laboratory,\\
Pine Street and Kirk Road, MS 357, \\
Batavia, Illinois, 50510\\
yinh@fnal.gov}

\maketitle

\begin{history}
\received{\today}
\revised{Day Month Year}
\end{history}

\begin{abstract}
We present the most recent precision electroweak measurements of single $W$ and $Z$ boson
cross section and properties from the LHC and Tevatron colliders, analyzing data collected by
ATLAS, CDF, CMS, D0, and LHCb detectors. The results include the measurement of the single
$W$ and $Z$ boson cross section at LHC, the differential cross section measurements, the measurement of 
$W$ boson mass, the measurement of $W$ and $Z$ charge asymmetry. 
These measurements provide precision tests on the electroweak theory, high order predictions 
and the information can be used to constraint parton distribution functions.
\end{abstract}


\section{Introduction}

The precision electroweak measurements provide stringent tests the on Standard Model (SM):
the measurement of single $W$ and $Z$ boson cross section
provides critical tests on the perturbative QCD and the higher order predictions,
the measurement of $W$ charge asymmetry can be used to constraint the parton
distribution functions (PDFs), and the measurement of $W$ mass and weak mixing
angle ($\sin^2\theta_W$) improves precision of the SM input parameters.
 
Recently, in both LHC and Tevatron, there are plenty of analyses related
to the single $W$ and $Z$ boson have been performed. 
The LHC is a $pp$ collider, results reviewed in this article used the collision data at
center-of-energy of 7 TeV and 8 TeV, collected by the ATLAS~\cite{atlas_det}, CMS~\cite{cms_det} 
and LHCb~\cite{lhcb_det} detectors.
The Tevatron is a $p\bar{p}$ collider, with center-of-energy of 1.96 TeV, using data collected
by the CDF~\cite{cdf_det} and D0~\cite{d0_det} detectors. The $W$ and $Z$ precision 
measurements have been performed based on these data.

At the Tevatron, the $W$ and $Z$ are produced with the $valence$ $quark$.
And at the LHC, in the productions of single $W$ and $Z$ boson, 
the $sea$ $quark$ and $gluon$ contributions become larger. 
Thus, single $W$ and $Z$ measurements at the Tevatron are complementary to that at the 
LHC. Furthermore, with knowing incoming $quark$
direction, the $Z$ forward-backward charge asymmetry at the Tevatron will be more sensitive
compared with that measurement at the LHC.

In this article, we review the $W$ and $Z$ cross section measurements from the LHC at first, then
review the measurements of boson transverse momentum
from both the Tevatron and LHC. In the end, the measurement of the $W$ boson mass,
$W$/$Z$ charge asymmetry measurements,
are presented. The physics results are collected from ATLAS, CDF, CMS, D0, and LHCb collaborations.

\section{The $W$ and $Z$ total cross section measurements}
Precise determination of the vector boson production cross section and their
ratios provide an important test of the SM.

\subsection{Inclusive cross section}
\begin{figure}[pb]
\centerline{\includegraphics[scale=0.3]{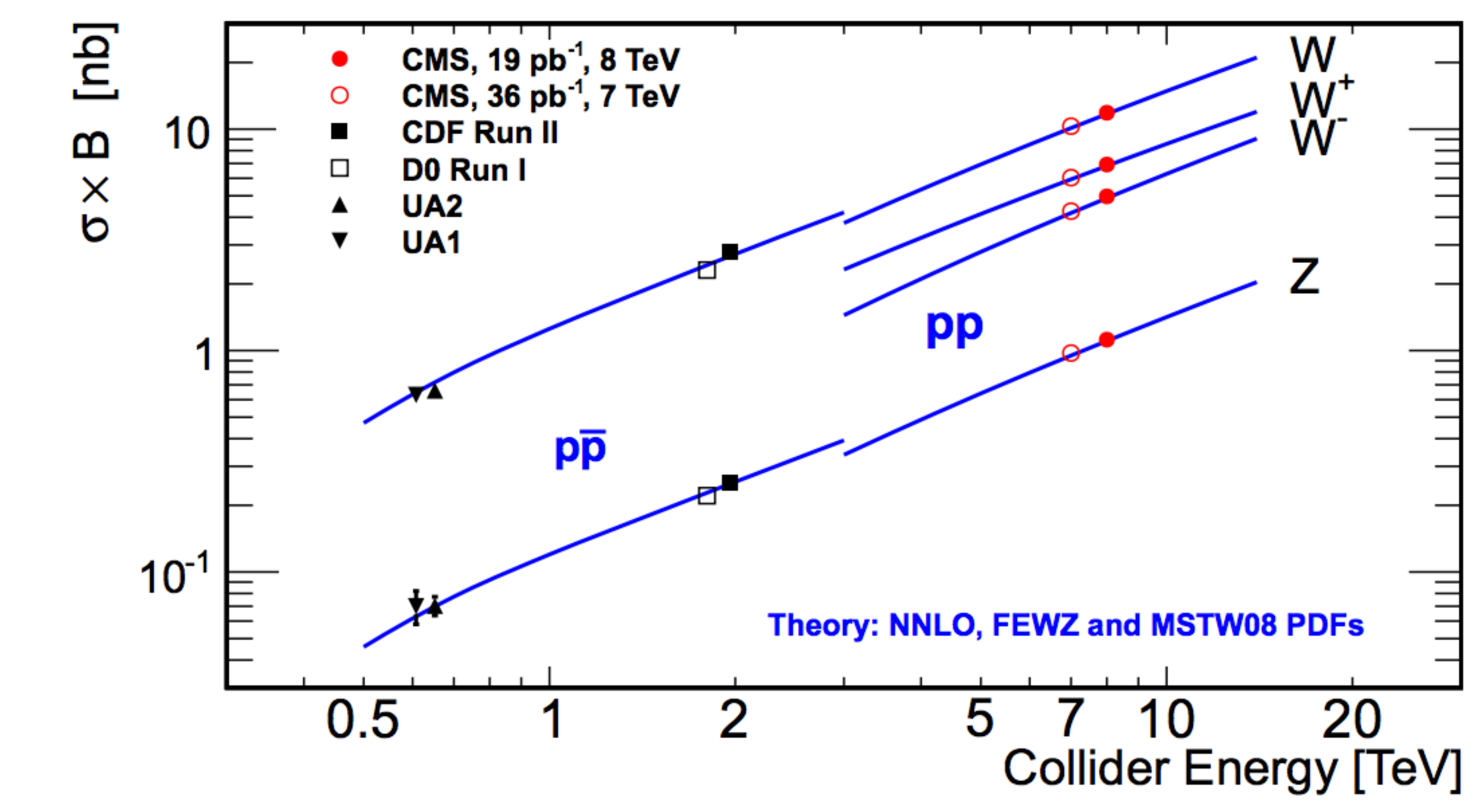}}
\vspace*{8pt}
\caption{The inclusive $W$ and $Z$ production cross section times branching ratios
as a function of center-of-mass energy measured from CMS and other lower-energy
colliders. The blue lines represent the NNLO theory predictions.. \label{cms_xsection_history}}
\end{figure}

At the LHC, the production of vector boson requires at least one sea quark,
and given the high scale of this process, the gluon contribution becomes significant.
Furthermore, the inclusive cross section of vector boson production is also sensitive to
the PDFs, the cross section predictions depend on the momentum distribution of the
gluon. And the theoretical uncertainty is another limitation for the cross section measurement,
the current available predictions are at next-to-leading (NLO) and next-to-next-to-leading
order (NLO) in perturbative QCD. In Fig.~\ref{cms_xsection_history}, the measured $W$ and $Z$
cross sections from different hadron colliders are compared with NNLO predictions, 
the predictions agree with measured value well.

Both ATLAS~\cite{atlas_tot_xsection} and CMS~\cite{cms_tot_xsection} collaborations presented the inclusive $W$ and $Z$ cross section
at $\sqrt{s}=7$ TeV, with electron and muon decay channels, using a data sample 
corresponding to 35 pb$^{-1}$ and 36 pb$^{-1}$ of integrated luminosity collected in 2010.
And furthermore, CMS~\cite{cms_tot_xsection1} using data corresponding to 19 pb$^{-1}$ of integrated luminosity collected 
in 2012 at $\sqrt{s} = 8$ TeV.

With $\sqrt{s}=7$ TeV data, the measured inclusive cross sections from CMS are $\sigma(pp\rightarrow WX)\times B(W\rightarrow l \nu) = 10.30 \pm0.02 (stat.) \pm 0.10 (syst.) \pm 0.10 (th.) \pm 0.41 (lumi.)$ nb and $\sigma(pp\rightarrow ZX)\times B(Z\rightarrow l^+ l^-) = 0.974 \pm0.007 (stat.) \pm 0.007 (syst.) \pm 0.018 (th.) \pm 0.039 (lumi.)$ nb,
and from ATLAS are $\sigma(pp\rightarrow WX)\times B(W\rightarrow l \nu) = 10.207 \pm0.021 (stat.) \pm 0.121 (syst.) \pm 0.347 (th.) \pm 0.164 (lumi.)$ nb and $\sigma(pp\rightarrow ZX)\times B(Z\rightarrow l^+ l^-) = 0.937 \pm0.006 (stat.) \pm 0.009 (syst.) \pm 0.032 (th.) \pm 0.016 (lumi.)$ nb.

\begin{figure}[pb]
\centerline{\includegraphics[scale=0.3]{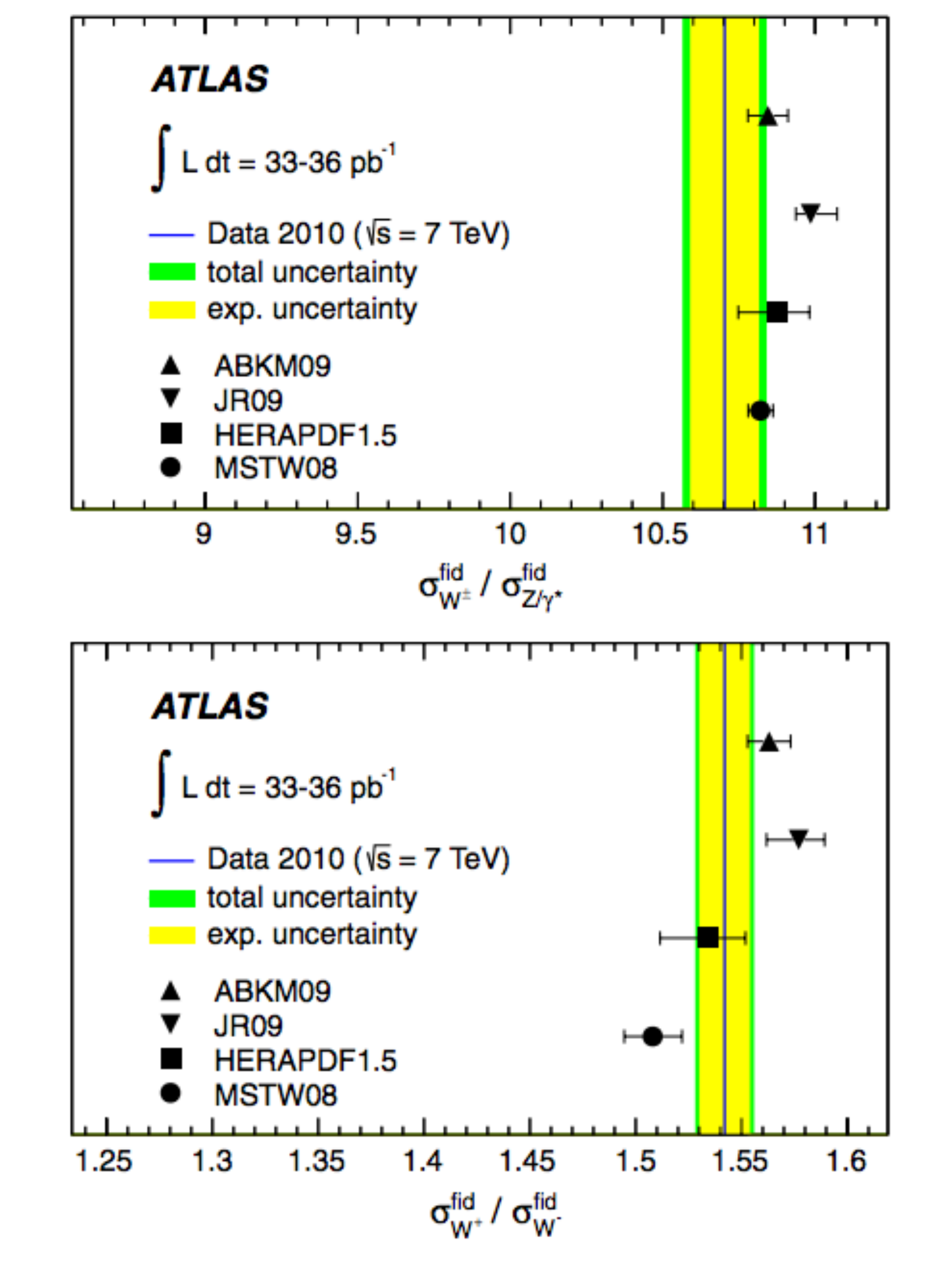}} 
\vspace*{8pt}
\caption{ The ratios between the measured inclusive cross section of $W^{\pm}$ and $Z$ (top panel),
and ratio between the measured inclusive cross section of $W^{+}$ and $W^{-}$ (bottom panel).
The inner bands represent experimental uncertainty, and outer bands represent total uncertainty.
The PDFs uncertainties from the {\sc abkm}, {\sc jr}, and {\sc mstw} predictions are considered in 68\% C.L.
The HERAPDF predictions include all three sources of uncertainty of that set.\label{atlas_xsection_wzratio}}
\end{figure}

\begin{figure}
\centerline{\includegraphics[scale=0.25]{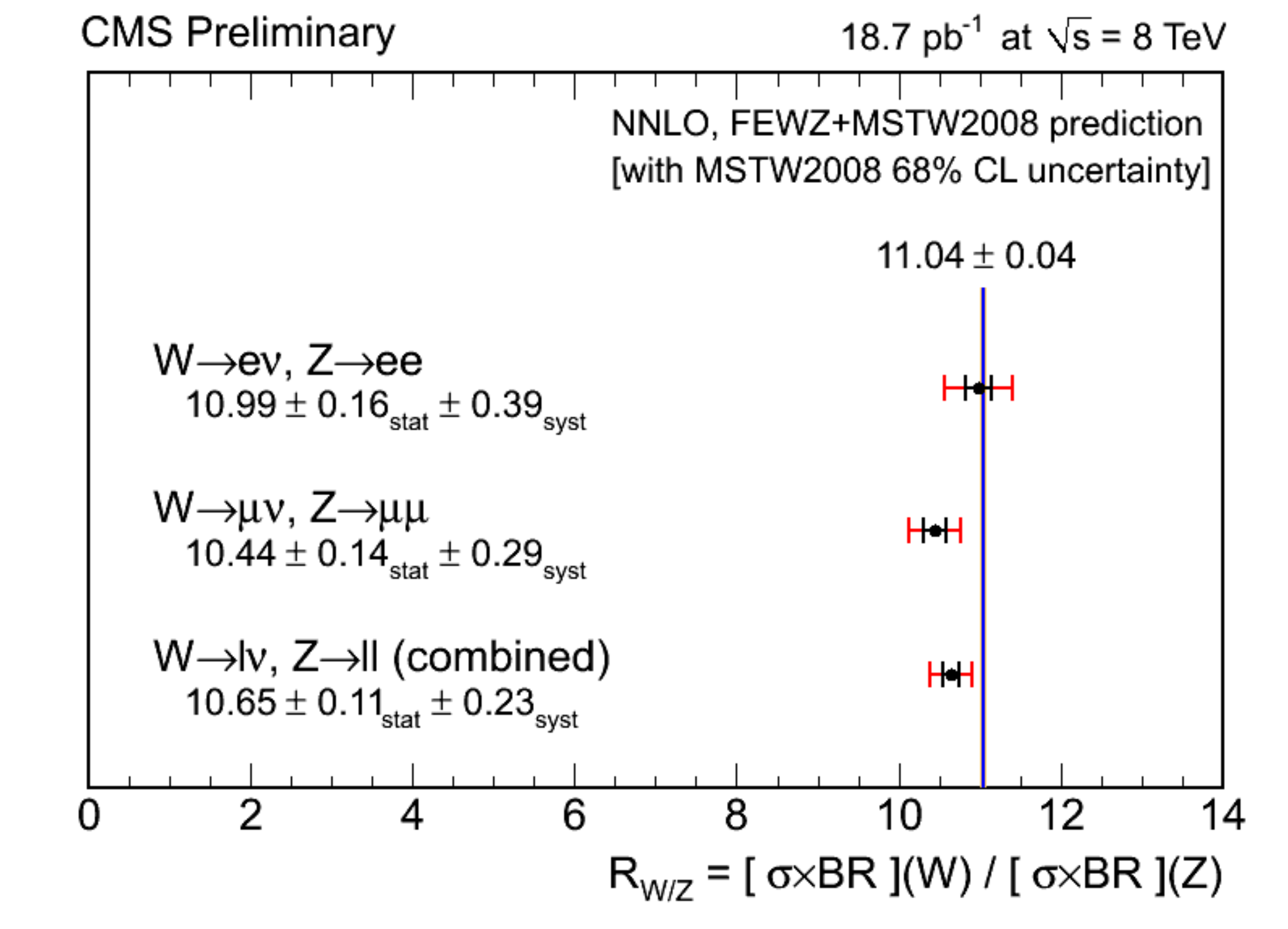}} 
\centerline{\includegraphics[scale=0.25]{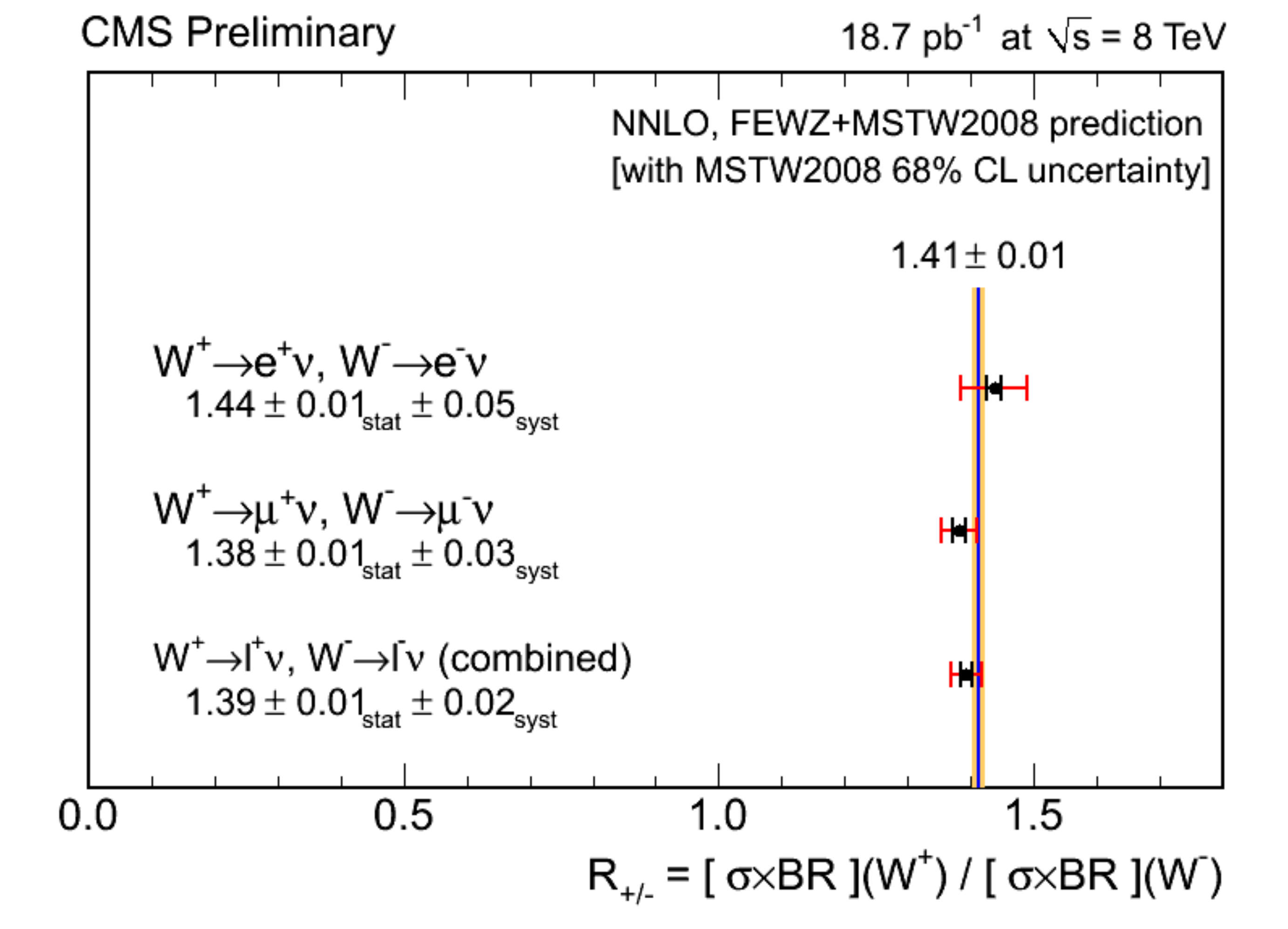}}
\vspace*{8pt}
\caption{The ratios between the measured inclusive cross section of $W^{\pm}$ and $Z$ (top panel),
and ratio between the measured inclusive cross section of $W^{+}$ and $W^{-}$ (bottom panel).
The black bars represent the statistical uncertainties, while the red errors bards include systematic
uncertainties.
Measurements in the electron and muon channels, and combined, are compared
to the theoretical predictions using {\sc fewz} and the {\sc mstw2008} PDF set. 
\label{cms_xsection_wzratio}}
\end{figure}

By presenting the measured cross section as a ratio will result a cancellation of systematic uncertainty,
including the luminosity uncertainty.
Thus, the ratios between measured $W^{\pm}$ and $Z$, also the ratio between $W^+$ and $W^-$,
are shown in Fig.~\ref{atlas_xsection_wzratio} and Fig.~\ref{cms_xsection_wzratio}.

The measurements of vector boson production cross section are consistent between
the electron and muon channels, and in agreement with NNLO order predictions.

\subsection{Differential cross section}
\begin{figure}
\centerline{\includegraphics[scale=0.3]{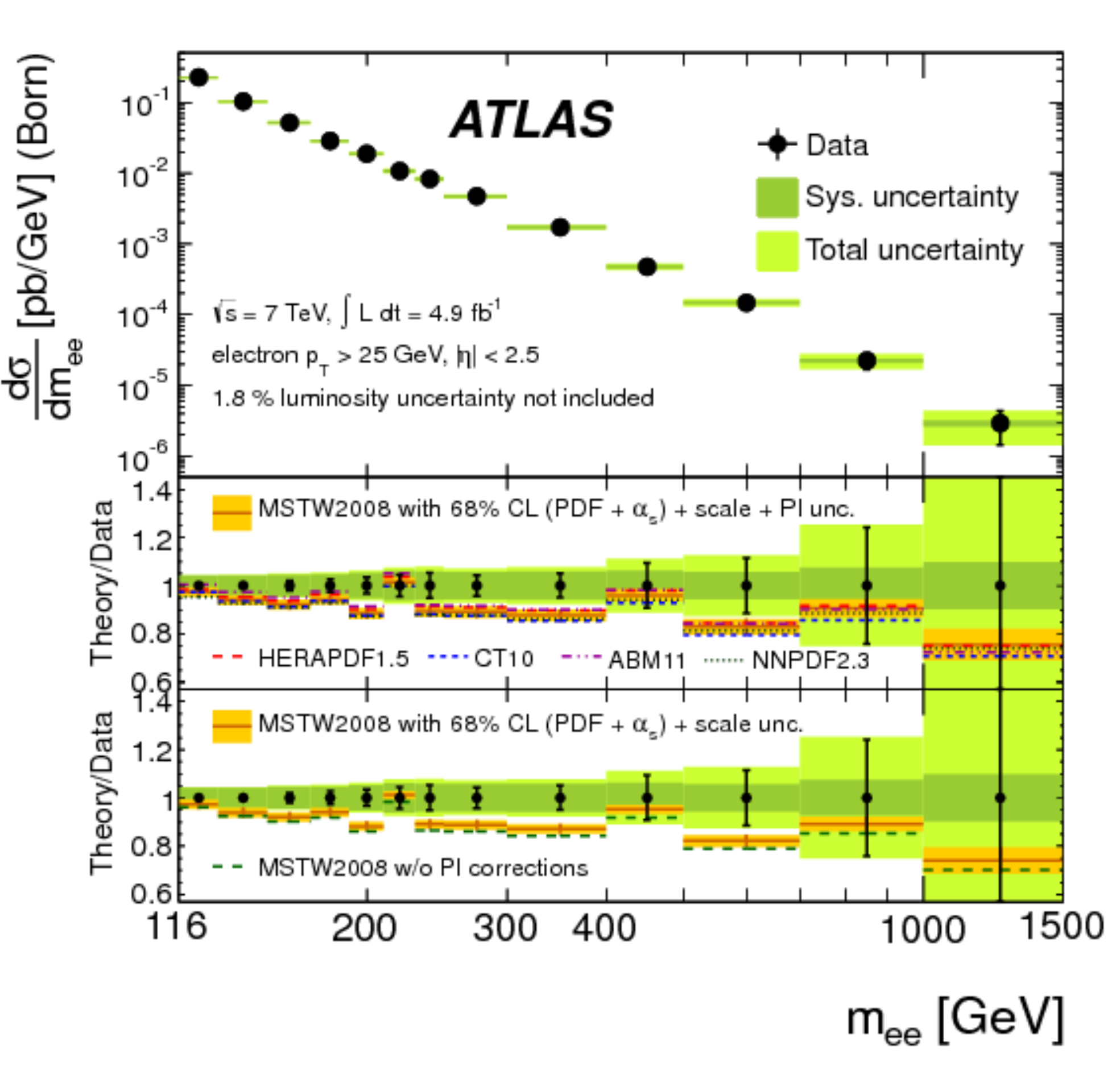}} 
\centerline{\includegraphics[scale=0.3]{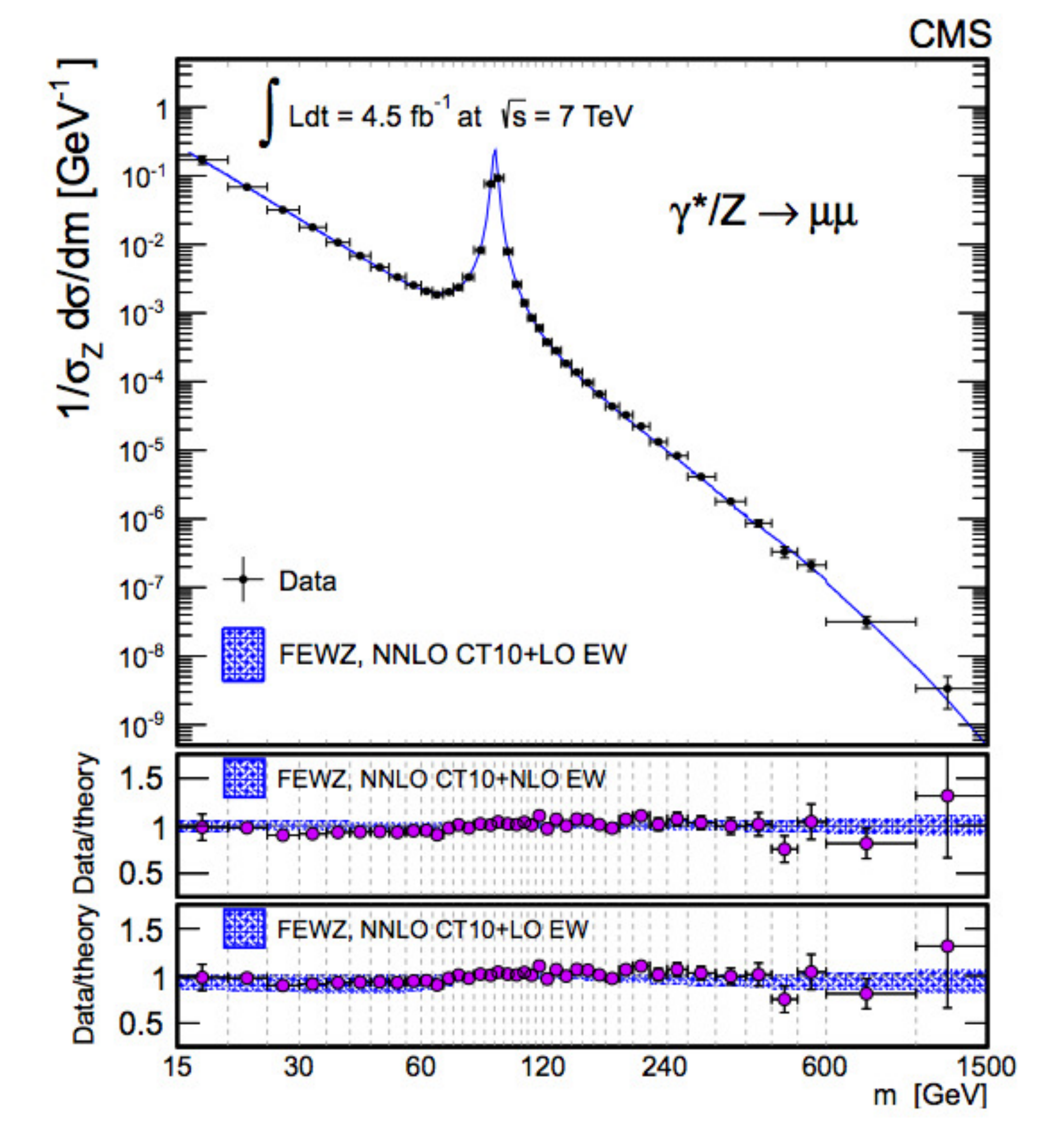}}
\vspace*{8pt}
\caption{Measured differential cross section as a function of invariant mass,
from ATLAS (top panel) and from CMS (bottom panel). 
The measured differential cross sections are compared with plenty of
high order predictions with different PDF sets. \label{lhc_xsection_diff}}
\end{figure}

The differential cross section of Drell-Yan (DY) process as a function of two
leptons invariant mass is sensitive to the PDFs, particular in the high mass region
(corresponding to the distribution function of antiquarks with high $x$ value),
and also provide stringent test on the high order perturbative QCD calculations.
Additionally, this process is a important source of background for other beyond SM
searches, the mass spectrum could be changed by new physics phenomena.
The ATLAS~\cite{atlas_diff_xsection} and CMS~\cite{cms_diff_xsection} report measurement 
of the high mass DY differential cross section measurement, as shown in 
Fig.~\ref{lhc_xsection_diff}. The measured differential cross sections
are constant with different NLO (or NNLO) theoretical predictions, with more data in the future,
this measurement could be used to constraint PDFs, in particular for antiquarks
with large $x$.

\begin{figure}
\centerline{\includegraphics[scale=0.3]{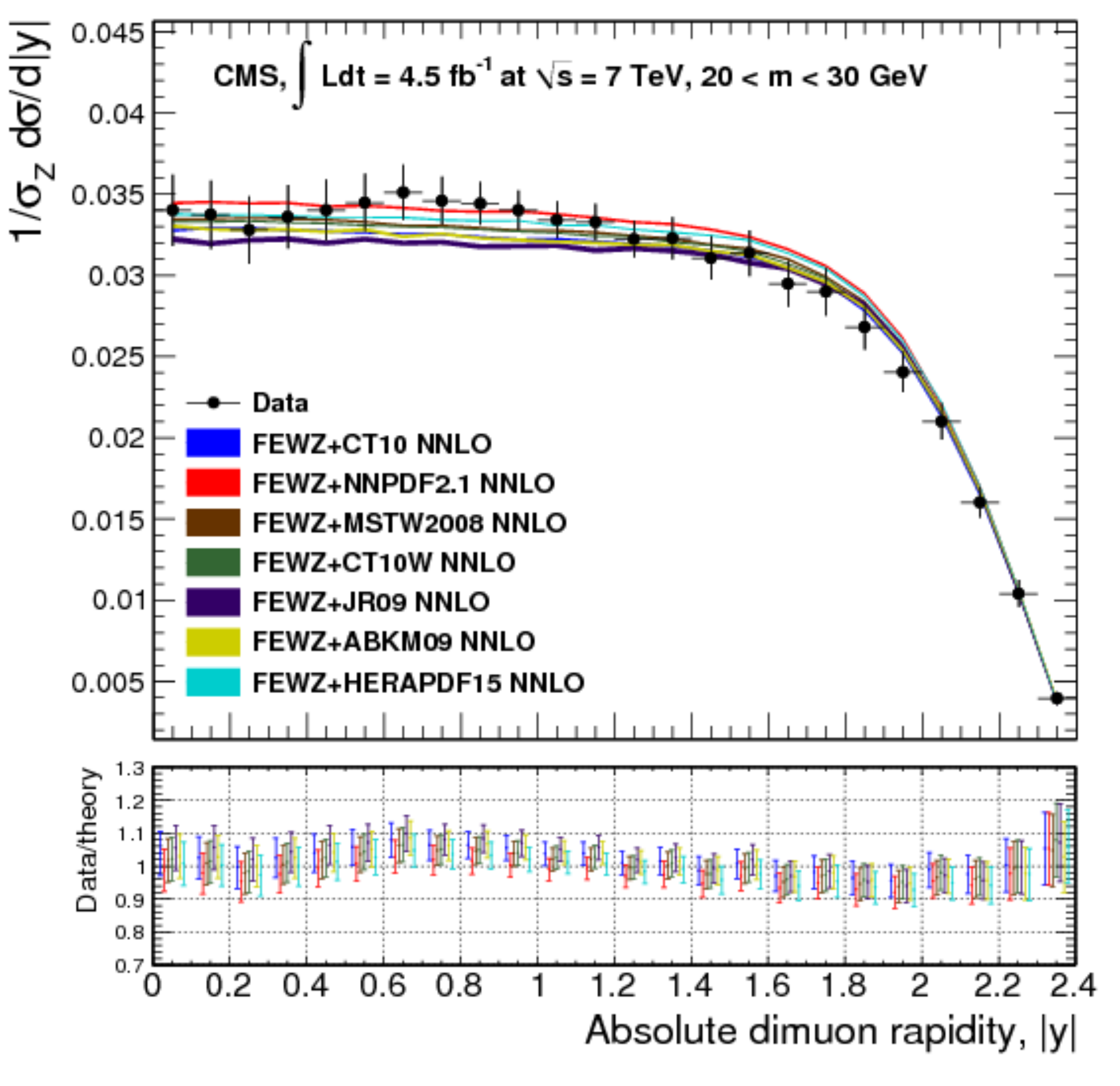}} 
\vspace*{8pt}
\caption{The measured differential cross section as a function of boson $y$,
in the mass region of 20-30 GeV. The uncertainty bands in the theoretical predictions
represent the statistical uncertainty. The bottom plot shows the ratio of
measured value to theoretical predictions.\label{cms_xsection_diff_2}}
\end{figure}

CMS also performed a double-differential cross sections measurement~\cite{cms_diff_xsection},
as functions of boson rapidity ($y$) and invariant mass.
By including additional boson rapidity information ($y$), the differential
cross section provides more information for the PDFs, while the 
low mass region, the high order effects and final state radiation (FSR)
become particularly important. The measured differential cross section
as a function of boson $y$ in the mass region of 20 to 30 GeV is shown
in Fig.~\ref{cms_xsection_diff_2}. The double -differential cross section 
measurement of DY will provides precise inputs to the future PDF sets fitting.

\subsection{Lepton universality}
\begin{figure}
\centerline{\includegraphics[scale=0.3]{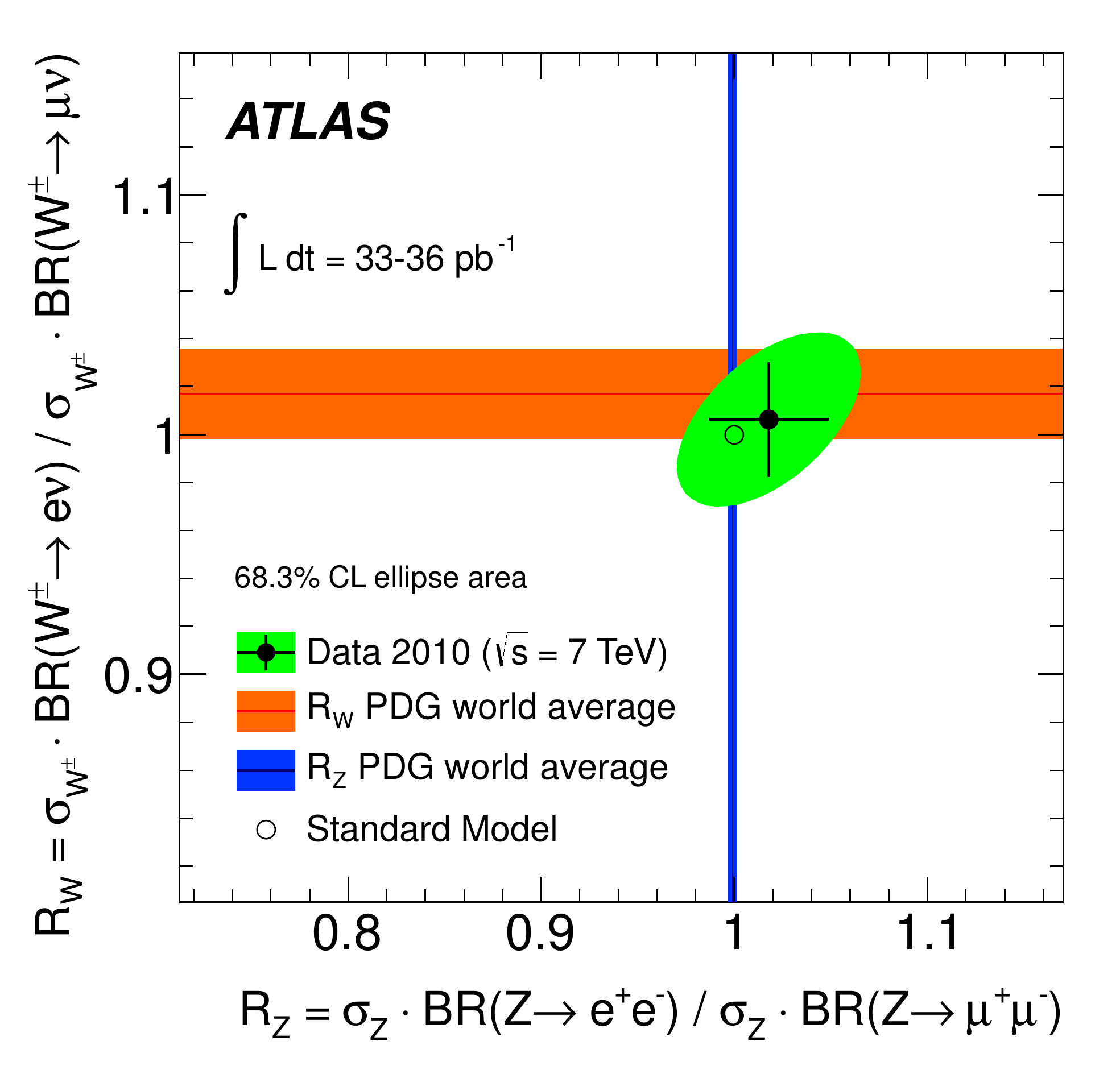}} 
\vspace*{8pt}
\caption{The correlated measurement of the electron to muon cross section ratios
in $W$ and $Z$ events. The vertical band represents the uncertainty from the $Z$ measurement,
and the horizontal band represents the uncertainty from the $W$ measurement.
The contour illustrates the 68\% C.L. for the correlation between $W$ and $Z$
measurements. \label{atlas_xsection_lepton}}
\end{figure}

The coupling of the leptons ($e$, $\mu$, $\tau$) to the gauge vector boson
are independent. The ATLAS performed a measurement of lepton universality~\cite{atlas_tot_xsection},
by comparing the cross sections measured using electron channel and muon
channel. The ratio between electron and muon channels results is shown
in Fig.~\ref{atlas_xsection_lepton}. The result confirms $e$-$\mu$ universality in
both $W$ and $Z$ decays.

\subsection{$W$ and $Z$ cross section measurements at LHCb}
\begin{figure}
\centerline{\includegraphics[scale=0.3]{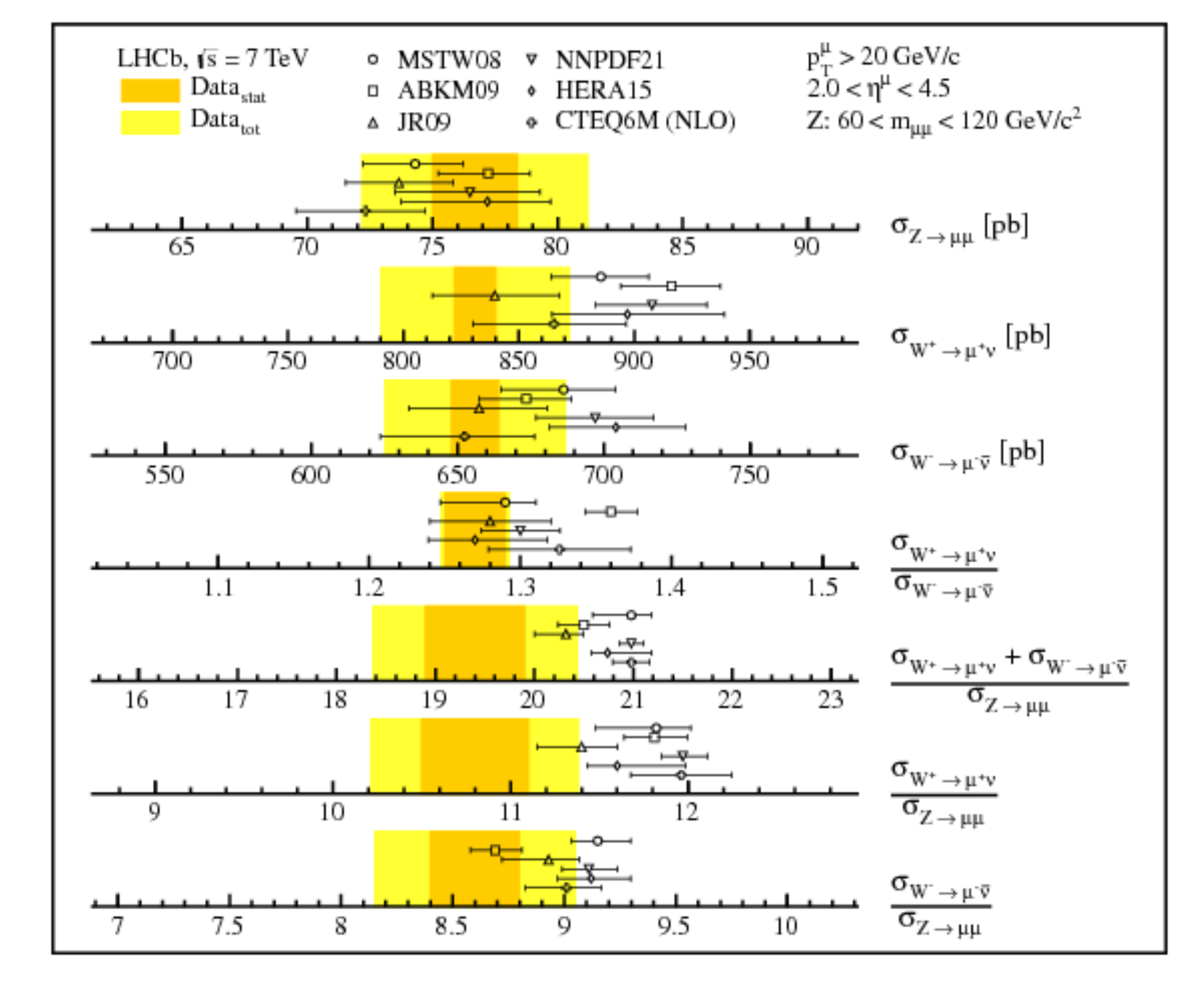}} 
\vspace*{8pt}
\caption{Measurements of the $Z$, $W^+$ and $W^{-}$ cross-section and ratios from LHCb, 
data are shown as bands which the statistical (dark shaded/orange) and total 
(light hatched/yellow) errors. The measurements are compared to NNLO and NLO 
predictions with different PDF sets for the proton, shown as points with error bars. 
The PDF uncertainty, evaluated at the 68\% confidence level, and the theoretical 
uncertainties are added in quadrature to obtain the uncertainties of the predictions.\label{lhcb_xsection_wzratio}}
\end{figure}

\begin{figure}
\centerline{\includegraphics[scale=0.25]{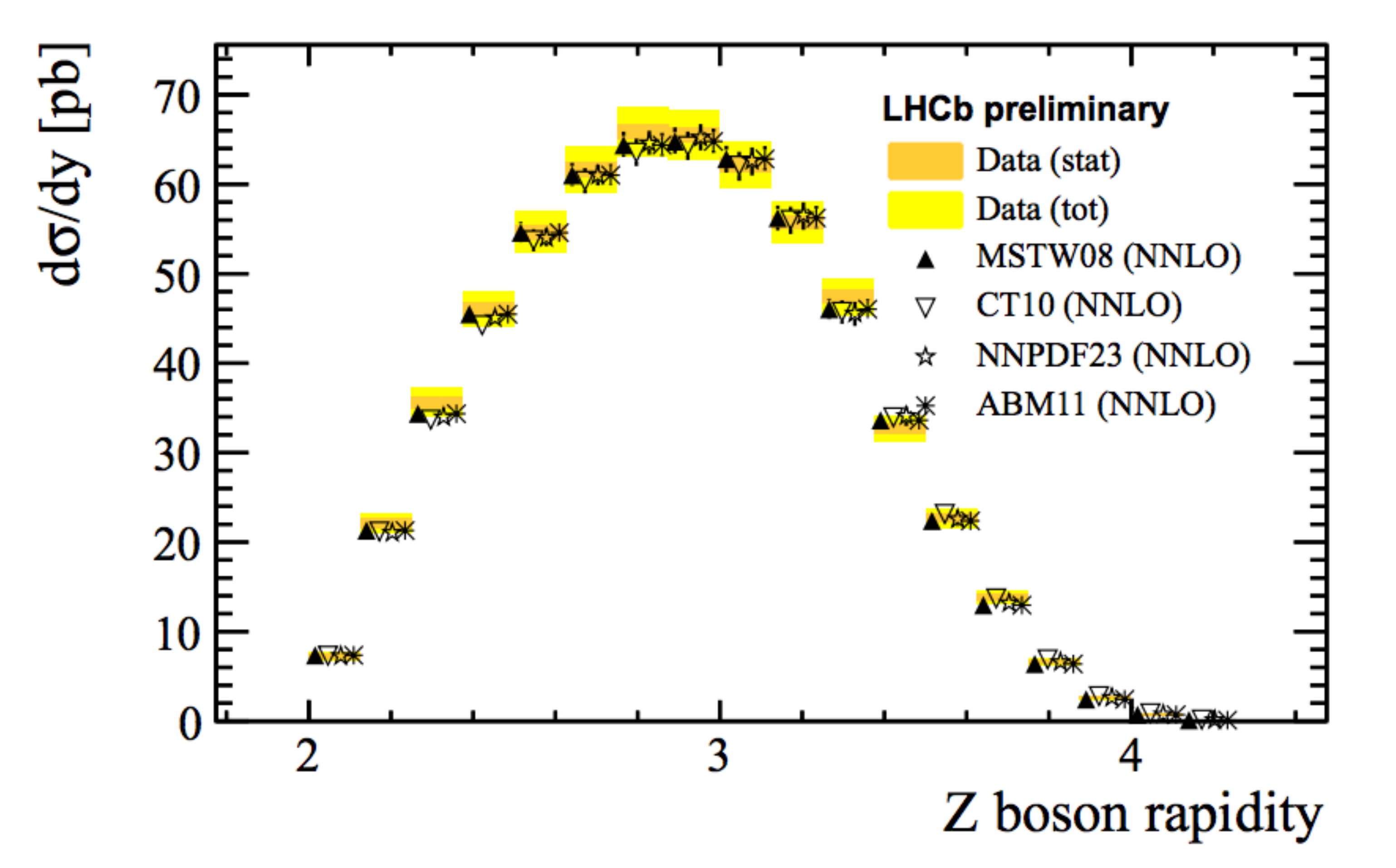}} 
\vspace*{8pt}
\caption{Differential cross section measurement as a function of $Z$ rapidity at LHCb. 
The yellow band represent the measured data value, while the NNLO
predictions are shown as points with error bars reflecting the uncertainties. \label{lhcb_xsection_diff}}
\end{figure}

In the LHCb, the measurement of the $W$ and $Z$ cross section
provide a unique test on the SM with high rapidity $W$ and $Z$ events.
Thus, the LHCb results have better precision than ATLAS and CMS in
the forward region ($2.0<\eta<5.0$). The most recent results~\cite{lhcb_tot_xsection,lhcb_zee_xsection,lhcb_zll_xsection,lhcb_diff_xsection}
are for the inclusive cross section measurements, and the differential cross
section measurement in different channels, as functions of rapidity or invariant mass.
The LHCb inclusive cross section measurements results are shown in Fig.~\ref{lhcb_xsection_wzratio},
and one of the differential cross section measurement from LHCb is shown in Fig.~\ref{lhcb_xsection_diff}.

\section{The boson transverse momentum measurements}
Initial state QCD radiation from the colliding
parton can change the kinematic of the Drell-Yan process system, 
results $Z$ boson be boosted with a transverse momentum, thus the precision measurement
of $Z$ boson $p_T$ can provide a stringent test on the higher order
QCD perturbative calculation. And this measurement can be used
to reduce the theory uncertainty of the measurement of $W$ mass .
In the low $p_T$ region, this process is dominated by soft and
collinear gluon emission, with the limitation of standard perturbative
calculation, the QCD resummation methods are used in the low $p_T$ region. While in the high 
$p_T$ region, the process is dominated by single parton emission,
thus, fixed-order perturbative calculations are used in the high $p_T$ region.

\subsection{Traditional method}
\begin{figure}
\centerline{\includegraphics[scale=0.3]{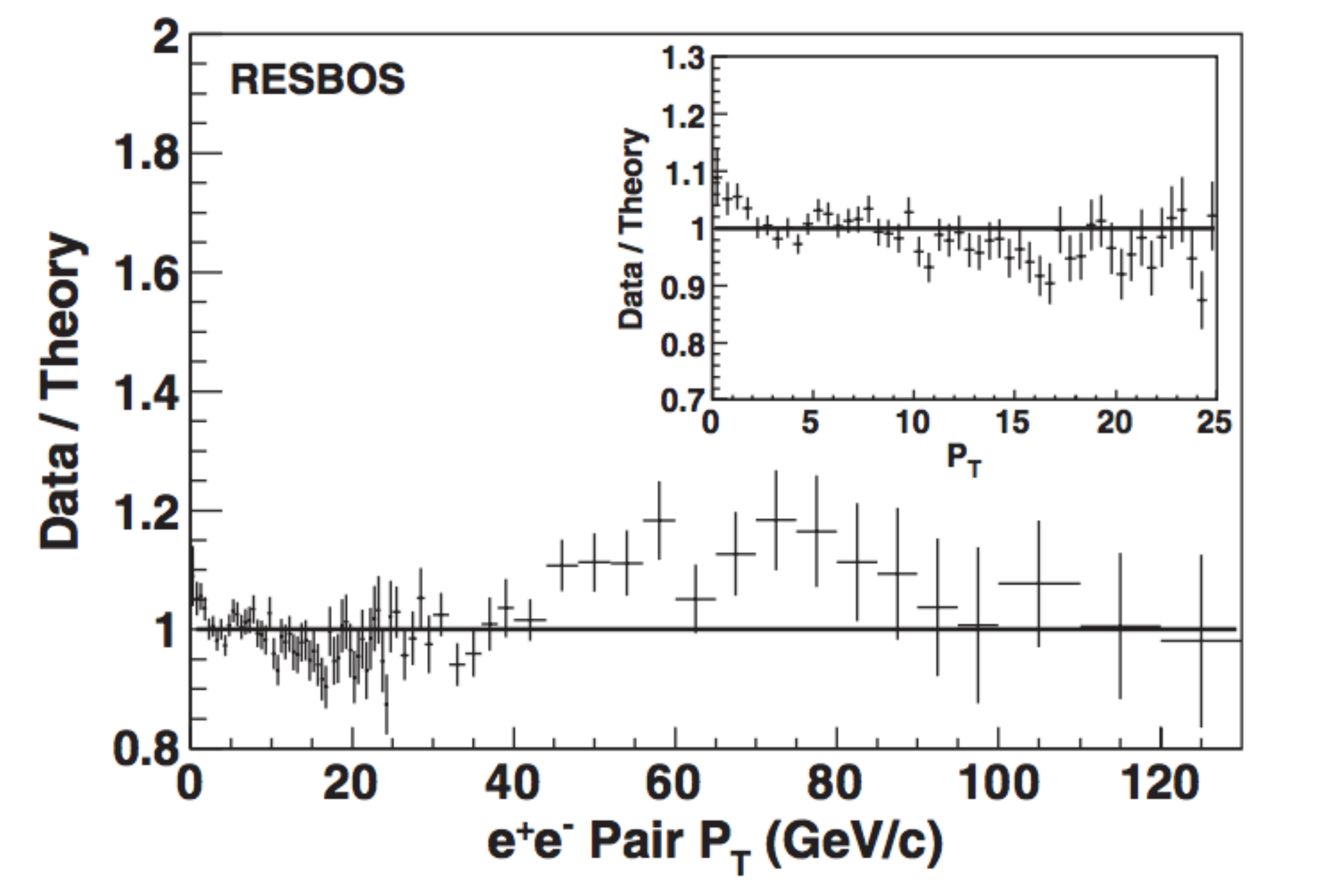}} 
\vspace*{8pt}
\caption{The ratio of the measured cross section to the {\sc resbos}
prediction in the $p_T < $ 130 GeV/$c$ region. The {\sc resbos} total
cross section is normalized to the data. The insert is an expansion of the low
$p_T$ region. \label{cdf_zptp}}
\end{figure}

\begin{figure}
\centerline{\includegraphics[scale=0.3]{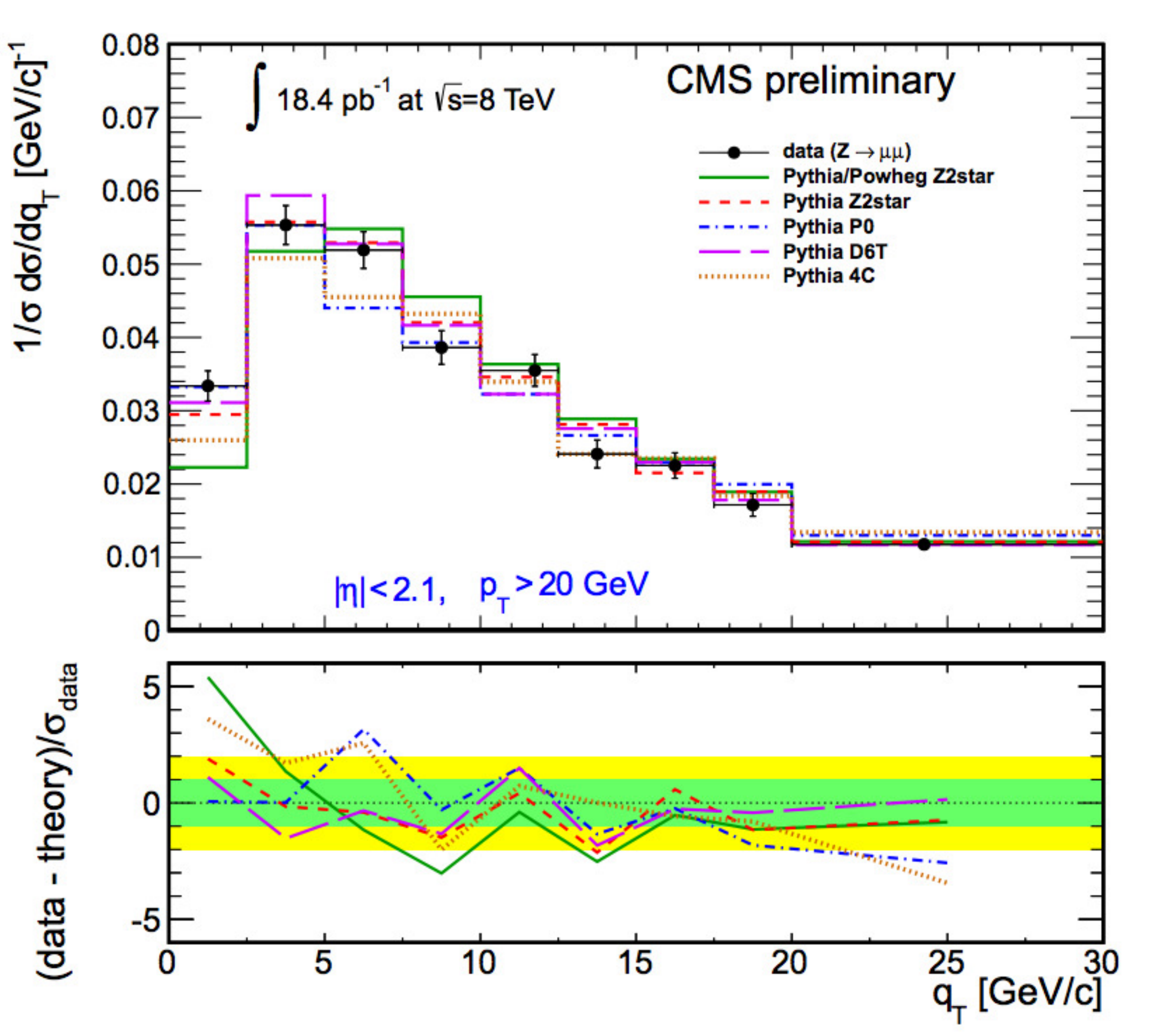}} 
\vspace*{8pt}
\caption{The transverse momentum distribution of $Z$ boson measurement from CMS.
The measure values (black points) are compared with different predictions. 
In the bottom portion, the difference between data and prediction decided by
the uncertainty of data is shown, with the green (inner)
and yellow (outer) bands represent the 60\% and 90\% C.L. experimental uncertainties. \label{cms_zpt}}
\end{figure}

In general, the boson $p_T$ can be directly measured by using the reconstructed boson
$p_T$.
As has been done before~\cite{d0_zpt},
CDF performed a precision
measurement of transverse momentum cross section of $e^+e^-$ pairs in the
$Z-$boson mass region of 66-116 GeV/$c^2$, with 2.1 fb$^{-1}$ of integrated luminosity~\cite{cdf_zpt}.
Fig.~\ref{cdf_zptp} shows the ratio between measured data value and {\sc resbos} predictions.
And CMS reported a similar measurement using data corresponding to 18.4 pb$^{-1}$ of integrated
luminosity~\cite{cms_zpt}, as shown in Fig.~\ref{cms_zpt}. The overall agreement 
with predictions from the SM is observed,
and the results have sufficient precision for the refinements of phenomenology in the future.

\subsection{A Novel method}

\begin{figure}[bt]
\centerline{\includegraphics[scale=0.4]{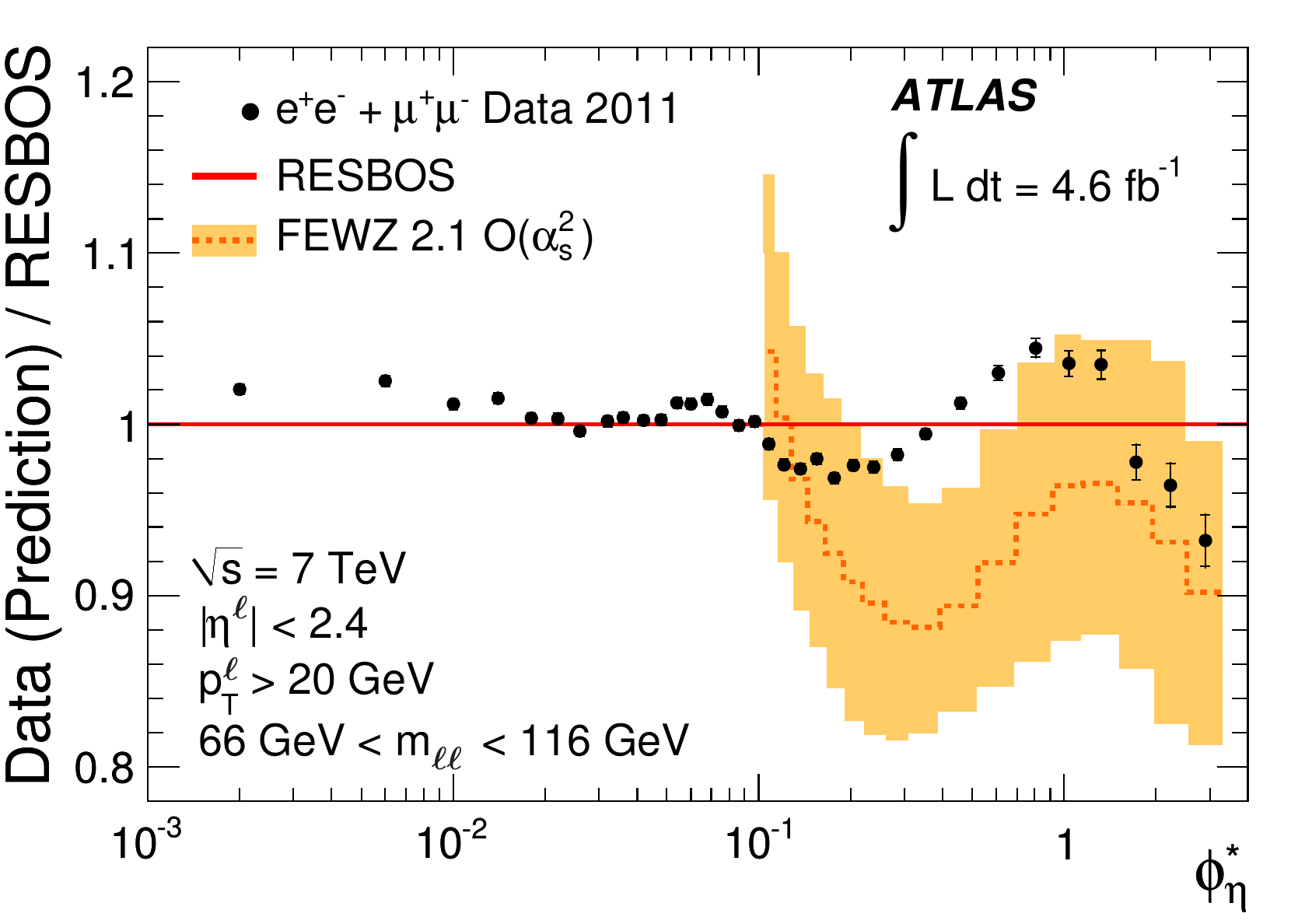}} 
\vspace*{8pt}
\caption{The ratio of the combined normalized differential cross section to 
{\sc resbos} predictions as a function of $\phi^*$. The inner and outer error bars on 
the data points represent the statistical and total uncertainties, respectively. 
The measurements are also compared to predictions, which are represented by a 
dashed line, from {\sc fewz} 2.1. Uncertainties associated to this calculation are 
represented by a shaded band. \label{atlas_zpt}}
\end{figure}

\begin{figure}
\centerline{\includegraphics[scale=0.4]{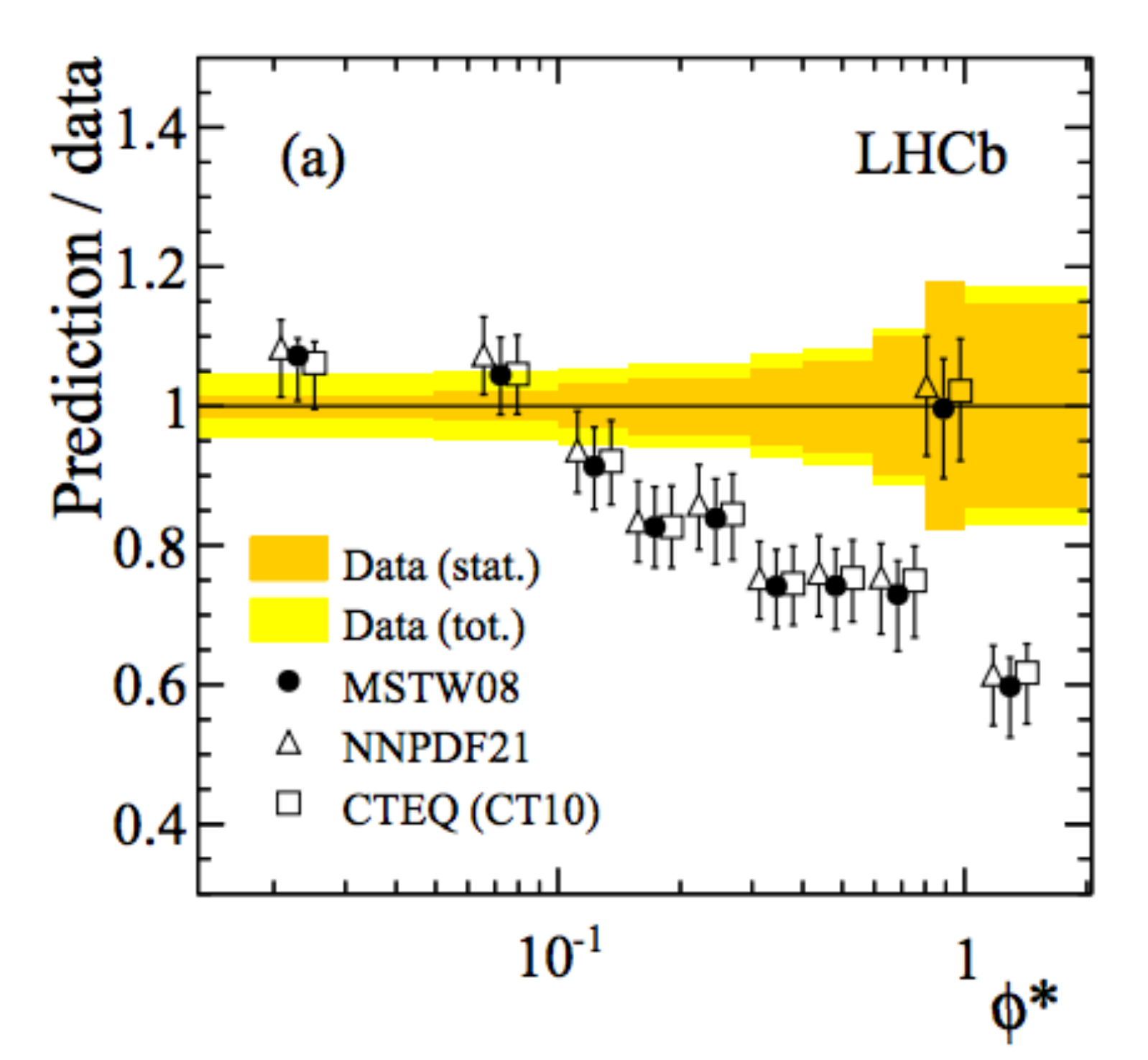}} 
\vspace*{8pt}
\caption{Ratio of various QCD predictions to the measured data values as a function
of $\phi^*$, using LHCb data. The NNLO QCD predictions to the measured data are shown
in black points. \label{lhcb_zphi}}
\end{figure}

In the previous $Z$ $p_T$ measurements, the uncertainty is dominated by
the experimental resolution and efficiency, and the choice of bin widths
is limited by the experimental resolution rather than event statistics.
In order to improve the precision of the boson $p_T$ measurements, 
D0 published measurement of $Z$ boson $p_T$~\cite{d0_zphi}, presented with a novel variable ($\phi*$)~\cite{zphi_th}.
The ATLAS and LHCb performed same measurements of $\phi^*$~\cite{atlas_zpt,lhcb_zee_xsection},
as shown in Fig.~\ref{atlas_zpt} and Fig.~\ref{lhcb_zphi}. 
The measured results provide sufficient information to the future retuning of
theoretical predictions. 

\section{The $W$ and $Z$ properties measurements}
There are many other electroweak measurements from Tevatron and LHC, which 
measure input parameters of the SM with better precision, 
including the measurement of $W$ mass, the measurement of $W$
charge asymmetry, and the measurement of weak mixing angle. 

\subsection{Measurement of the $W$ boson mass}

\begin{figure}
\centerline{\includegraphics[scale=0.3]{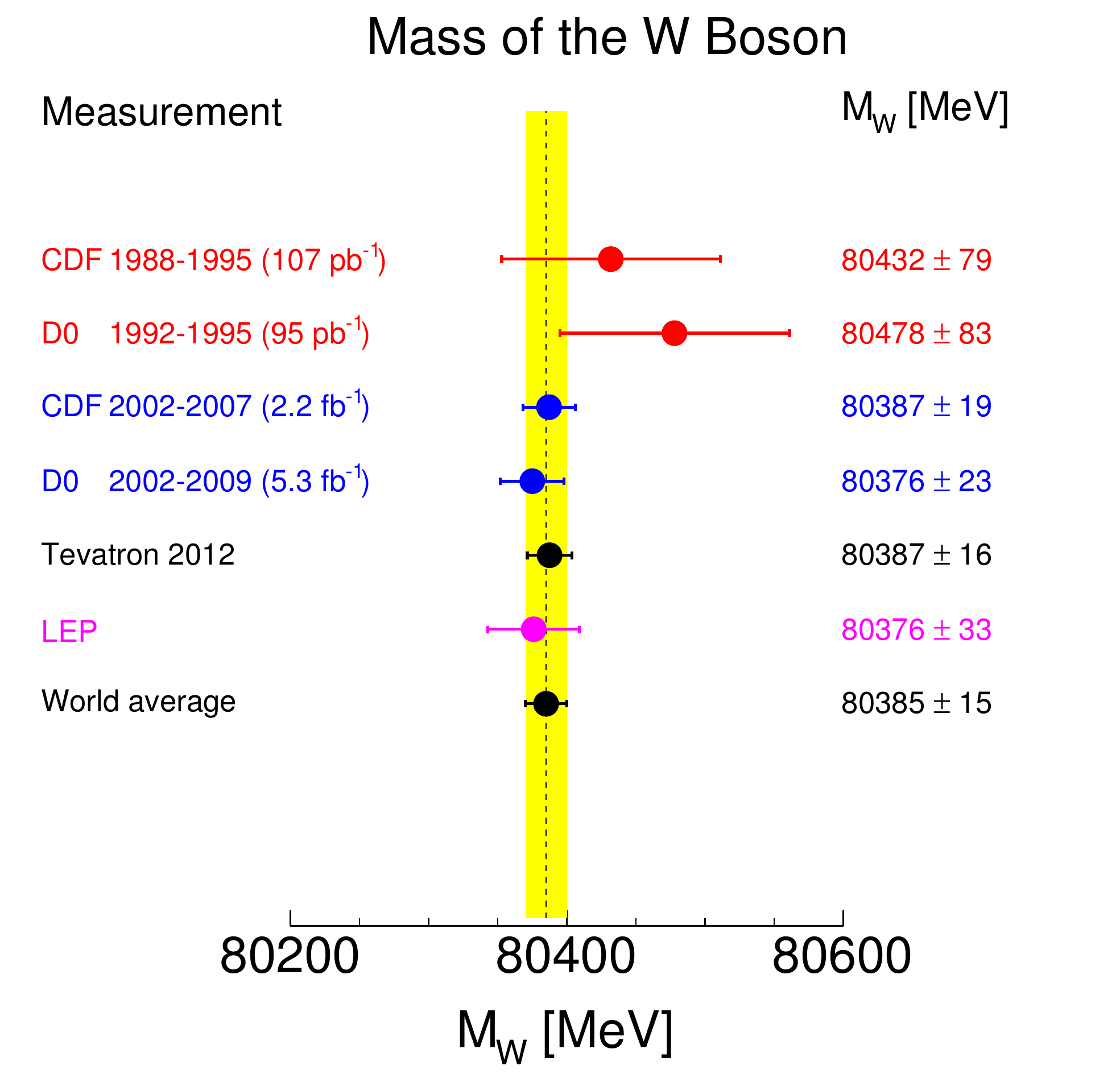}} 
\vspace*{8pt}
\caption{Measured $W$ boson mass from the CDF and D0 Run I (1989 to 1996)
and Run II (2001 to 2009), the new Tevatron average, and the LEP
combined results, and the world average obtained by combining the Tevatron
and LEP measurements, with assumption that no correlation between them.
The shaded band is the new world-average uncertainty (15 MeV). \label{amass}}
\end{figure}

The precision measurement of $W$ mass contribute to
the understanding of the electroweak interaction. 
By combining the more recent measurement of $W$ mass from
both CDF~\cite{cdf_wmass}, D0~\cite{d0_wmass}, and previous
Tevatron measurements, 
the Tevatron average value is $80387\pm16$ MeV~\cite{wmass}.
The combination of $W$ mass with the LEP average further
reduces the uncertainty to 15 MeV, as shown in Fig.~\ref{amass}.

\subsection{$W$ charge asymmetry measurements}
The $W$ charge asymmetry is sensitive to PDFs. The Tevatron is a $p\bar{p}$ collider,
the $u$ quark tends to carry higher momentum than $d$ quark, thus arise a
non-zero asymmetry as a function of $W$ rapidity. The LHC is $pp$ collider,
proton has two valance $u$ quarks and one valence $d$ quark, there are
more $W^+$ than $W^-$. In hadron colliders, the neutrino escapes the detector
without producing any measurable signal, without the neutrino longitudinal momentum,
the $W$ charge asymmetry can be measured as lepton charge asymmetry, which 
is a convolution of the $W$ boson asymmetry and $W$ $V$-$A$ decay.

\begin{figure}
\centerline{\includegraphics[scale=0.25]{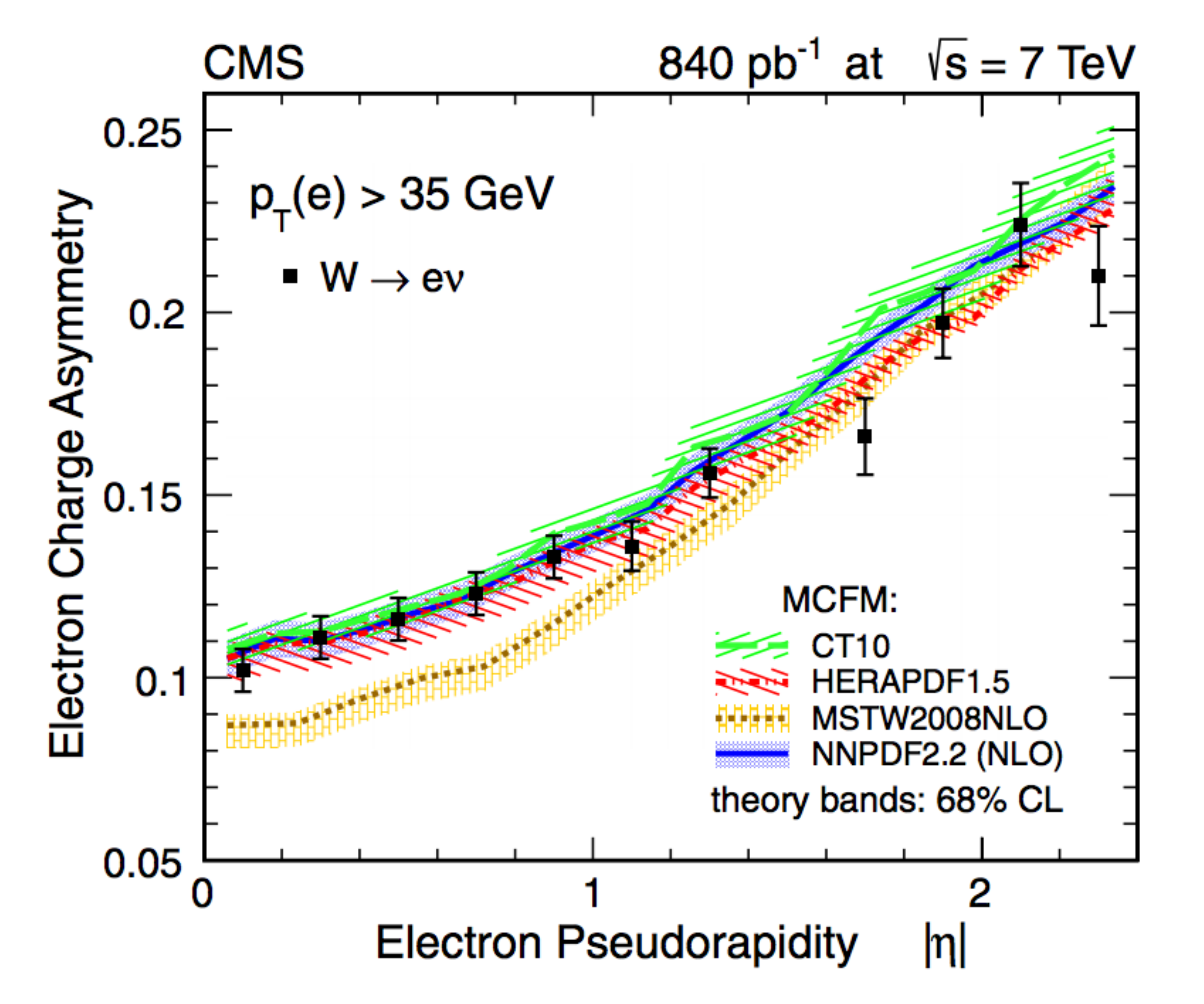}} 
\vspace*{8pt}
\caption{Lepton charge asymmetry measurement from CMS, using electron channel.
With a cut on the electron transverse momentum (35 GeV). The error
bars include both statistical and systematic uncertainties. \label{cms_wasymp}}
\end{figure}

\begin{figure}
\centerline{\includegraphics[scale=0.3]{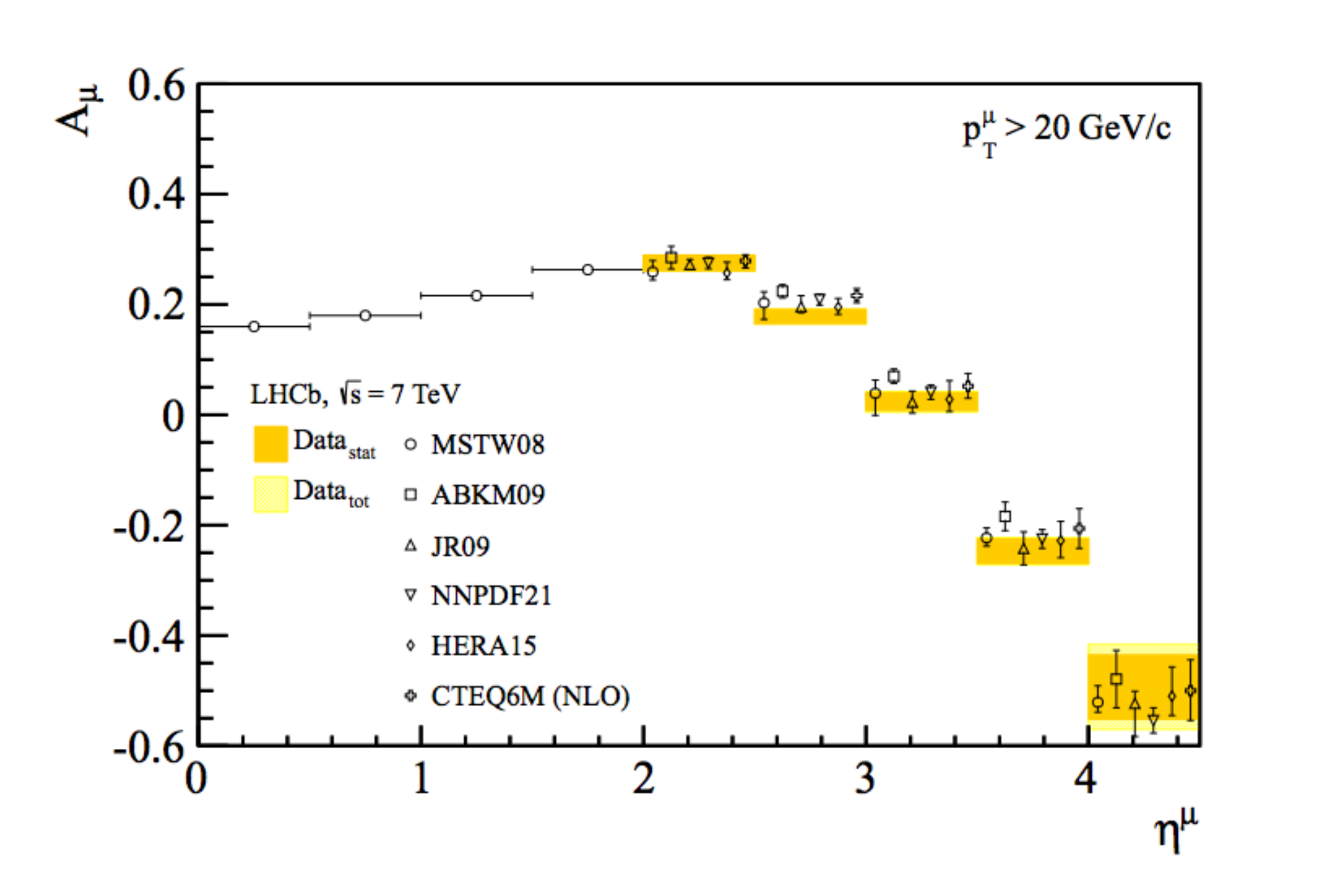}} 
\vspace*{8pt}
\caption{Lepton charge asymmetry measurement from LHCb. The
orange bands correspond to the statistical uncertainties, the yellow bands
represent total uncertainty. The measured values are compared with different
NNLO predictions. \label{lhcb_wasymp}}
\end{figure}

The lepton asymmetry has also been measured at the LHC in $pp$ 
collisions by the ATLAS~\cite{atlas_wasym} and CMS~\cite{cms_wasym}, \cite{cms_wasym2}
Collaborations using data corresponding to 31 pb$^{-1}$ and 840 pb$^{-1}$ of integrated luminosity, respectively.
One of CMS results is shown in Fig.~\ref{cms_wasymp}.
With 37 pb$^{-1}$ of integrated luminosity collected with LHCb detector, the lepton charge 
asymmetry is performed using muon channel~\cite{lhcb_tot_xsection}, as shown in Fig.~\ref{lhcb_wasymp}.
These measurements provide useful information for future PDF set fitting.

\subsection{Weak mixing angle measurements}

Weak mixing angle is one of fundamental parameters in the SM. It is one of key input parameters
related to the electroweak couples, for both charge current ($W$)
and neutrino current ($Z$). The weak mixing angle is a running
parameter in a wide region of center-of-energy, there are many measurements
have been done at the low energy experiments, like atomic parity violation~\cite{APV}, 
Milloer scattering~\cite{milloer}, and NuTeV~\cite{nutev}. In the $Z$ peak
region, the most precision measurements are from LEP $b$ quark asymmetries,
and SLD Left-Right hand asymmetries ($A_{LR}$)~\cite{lep}. The results from 
LEP and SLD are deviated by three standard deviations in different directions.

Recently at the hadronic collider, CDF~\cite{cdf_stw_old}, D0~\cite{dzero_1fb,dzero_5fb}, and 
CMS~\cite{cms_stw,cms_stw2} have been performed this measurement,
which show reasonable agreement with world average value. 
Due to the limitations from PDFs
and quark fragments, the dominated systematic uncertainty comes from
PDFs, which can be possibly suppressed by the update of PDFs sets.

\begin{figure}
\centerline{\includegraphics[scale=0.4]{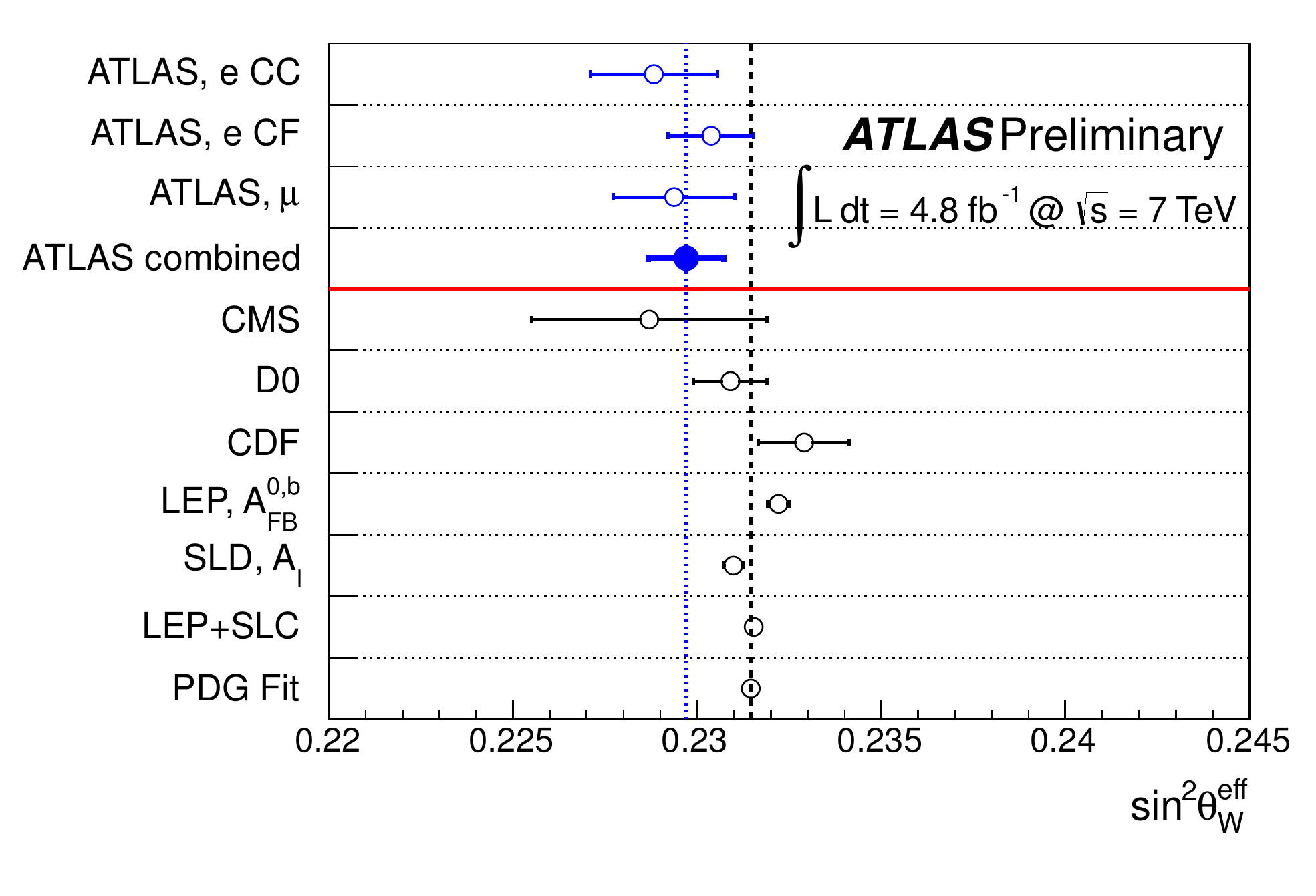}} 
\vspace*{8pt}
\caption{Comparisons of measured weak mixing angle from different experiments. And the D0 result
includes the 90\% C.L. PDF uncertainty instead of 68\% C.L. PDF uncertainty in other measurements. \label{stw_summary}}
\end{figure}

With data corresponding to 2 fb$^{-1}$ of integrated luminosity, CDF performed a 
measurement of weak mixing angle~\cite{cdf_stw} using $Z\rightarrow ee$ events. 
Using data corresponding to 4.8 fb$^{-1}$ of integrated luminosity, ATLAS
presented a measurement of weak mixing angle~\cite{atlas_stw}, after combining both
electron and muon channels, the measured value at ATLAS is $0.2297 \pm 0.0004(stat.)\pm0.0009(syst.)$.
The summary for weak mixing angle measurement from different experiments
is shown in Fig.~\ref{stw_summary}.
The PDF uncertainty is the key element for this measurement at hadron colliders, with 
more data and new PDF set, the precision of weak mixing angle from hadron collider
may be comparable to that from LEP.

\section{Summary}
Precision measurements with single $W$ and $Z$ bosons events provide stringent tests
on the SM. With the data collected from the LHC and Tevatron, good consistency
between SM and data is observed, these measurements also provide more information
for PDF fitting, critical tests on the high order predictions, and more precisely SM
input parameters. With good understanding on the detector response, more precision electroweak
results with larger data set will come out soon from both LHC and Tevatron.



\section{References}


\end{document}